\documentclass[%
 aip,
 sd,%
 amsmath,amssymb,
 reprint,%
]{revtex4-1}

\usepackage{graphicx}
\usepackage{dcolumn}
\usepackage{bm}

\graphicspath{{Pic/}}
\begin{document}

\preprint{AIP/123-QED}

\title[Intra-annual Principal Modes and Evolution Mechanism of the El Ni$\tilde{n}$o/Southern Oscillation]{Intra-annual Principal Modes and Evolution Mechanism of the El Ni$\tilde{n}$o/Southern Oscillation}


\author{Yongwen Zhang}
 \affiliation{Data Science Research Center, Faculty of Science, Kunming University of Science and Technology, Kunming 650500, China;} 
 \affiliation{Department of Physics, Bar-Ilan University, Ramat Gan 52900, Israel;}
\author{Jingfang Fan}%
 \affiliation{Potsdam Institute for Climate Impact Research, 14412 Potsdam, Germany;}
\author{Xiaoteng Li}%
 \affiliation{Harvest Found Management Co., Ltd, 100005 Beijing, China;}
 
\author{Wenqi Liu}
 \affiliation{Data Science Research Center, Faculty of Science, Kunming University of Science and Technology, Kunming 650500, China;} 
\author{Xiaosong Chen}
\email{chenxs@bnu.edu.cn}
\affiliation{School of Systems Science, Beijing Normal University, Beijing 100875, China}
%
\date{\today}

\date{\today}

\begin{abstract}
The El Ni$\tilde{n}$o-Southern Oscillation (ENSO) is one of the most important phenomena in climate. By studying the fluctuations of surface air temperature within one year between 1979-01-01 and 2016-12-31 of the region ($30^\circ$S-$30^\circ$N, $0^\circ$E-$360^\circ$E) with eigen-decomposition, we find that the temperature fluctuations are dominated by the two principal modes whose temporal evolutions respond significantly to the ENSO variability. According to introduce a ``micro-correlation'', we find that the coupling between the first principal mode and the temperature fluctuations in the El Ni$\tilde{n}$o region could result in different ENSO phases. Without this coupling, the El Ni$\tilde{n}$o region is in a normal phase. With the strong coupling between the El Ni$\tilde{n}$o region and the Northern Hemisphere, an El Ni$\tilde{n}$o event will appear with a high probability. Then this coupling changes to be strong between the El Ni$\tilde{n}$o region and the Southern Hemisphere accounting for the fast decay of El Ni$\tilde{n}$o after boreal winter, even leading to a La Ni$\tilde{n}$a event. Moreover, the coupling between the El Ni$\tilde{n}$o region and the second principal mode is found to be strong in normal or La Ni$\tilde{n}$a phases in response to the normal or strong Walker Circulations. We conjecture that the temporal evolutions of these couplings for the first and second principal modes are controlled by the meridional and zonal ocean-atmospheric circulations respectively.
\end{abstract}

\pacs{Valid PACS appear here}
\maketitle


\section{\label{sec:level1}Introduction}
The complexities of climate systems are generally due to the existence of multi-scales phenomena \cite{Hartmann1994Global}, which have a wide variety of space and time scales. Small-scale phenomena such as convection and precipitation \cite{Neelin2011Climate} assumed to be driven by large-scale forces such as extratropical cyclones, planetary-scale waves, and meridional circulations that extend over thousands of kilometers to transport energy between the tropics and the polar regions. The El Ni$\tilde{n}$o/Southern Oscillation (ENSO), referring to vary between anomalously warm (El Ni$\tilde{n}$o) and cold (La Ni$\tilde{n}$a) phases in eastern Pacific, can cause the global impacts on year-to-year time scales \cite{Mcphaden2006ENSO,Wang2017El} . In 1969, Bjerknes postulated that the Bjerknes feedback (the Walker circulation) was essential to the ENSO \cite{Bjerknes1969ATMOSPHERIC}. Still, the mechanism of
ENSO has not been fully understood \cite{Mcphaden2015Playing}. The alternation between warm and cold phases is quite irregular \cite{Timmermann2018}. There was controversy over whether the ENSO is controlled by stochastic processes \cite{Neelin1998ENSO}. Teleconnections are widely observed near the equator, since heat frequently exchanges between the ocean and atmosphere in the region, and energy is transported over a great distance from the region by the general circulations \cite{Jin2000An, Knox2017Atmospheric}. From the physics point of view, a small-scale fluctuation could trigger abrupt changes in large-scale, if the precondition is satisfied that the system exits long-range correlations (teleconnections). \cite{Stanley1973Introduction,Sornette2006,Zhang2018b}. Thus the ENSO could be triggered by small-scale and short-term processes. 

Although the ENSO is as well known as an inter-annual phenomenon, a significant intra-annual  property of the ENSO has been found that it is the synchronization with annual cycle--El Ni$\tilde{n}$o events tend to peak in boreal winter and fall rapidly in boreal spring \cite{Stein2011Phase,Mcgregor2011The}. To capture this synchronization and the rapid termination of El Ni$\tilde{n}$o events, the interaction between inter-annual and intra-annual timescales need to be considered \cite{Stuecker2013A,ren2016enso,Guilyardi2009Understanding}. The ENSO could be coupling with intra-annual oscillations \cite{Jin1994El, Tziperman1994El,stein2014enso,stuecker2015nino}, even with higher frequency processes \cite{Fedorov2010How,Kessler2002Is}. 

Eigen analysis has been a classic technique to distinguish multiple physical processes into a combination of some single processes and capture the individual feature. It is widely used in community detection, image recognition and empirical data analysis \cite{Newman2006,Zhang2019b,Hofmann2001Unsupervised}. It has also emerged as a popular tool in climate science \cite{hannachi2007empirical} and been used to search for the important features of the ENSO \cite{zhang1997enso,kawamura1994rotated,mcgregor2013meridional}. Inter-annual principal modes associated with the ENSO are found based on eigen analysis \cite{zhang1997enso, Stuecker2013A}. On the other hand, other methods such as climate networks have also been used to study and predict the El Ni$\tilde{n}$o in recent decades \cite{Donges2009b,backj2009,Yamasaki2008a,Ludescher2014,Fan2017,Meng2017,Zhang2019a,Zhang2018a}. Actually, the essential of climate networks depends on the correlation matrix as well as the eigen method\cite{Donges2015a,Wiedermann2016}. 

Here, we use eigen analysis to study the principal modes of daily surface air temperature. Note that we use the original temperature rather than abnormal temperature. Previous studies \cite{zhang1997enso,kawamura1994rotated,mcgregor2013meridional} mentioned above focused mainly on the inter-annual principal modes after removal of seasonality. Instead, we investigate the intra-annual principal modes including seasonality. According to eigen-decomposition for a correlation matrix in a 365 days' window, the principal modes will be obtained. The dynamics of the principal modes will be detected by the temporal evolution of the eigenvalues and eigenvectors. The properties of eigenvalues usually were neglected by the previous studies. From a physical perspective, eigenvalues play an important role to describe macroscopic properties in a physical system \cite{li2016critical,Hu2019}. Furthermore, the detailed spatial patterns are performed by the distributions of eigenvectors. This paper is organized as follows. Section \ref{sec:level2} describes the methods and data. In section \ref{sec:level3} we presented the results, and provides the conclusions in section \ref{sec:level4}.

\section{\label{sec:level2}Data and Methodology}
\subsection{\label{data}Data}

Our study is based on daily surface air temperature (2 m) and wind (10 m) provided by the European Centre for Medium-Range Weather Forecasts Interim Reanalysis (ERA-Interim) \cite{Dee2011The} on a $2.5^{\circ} \times 2.5^{\circ}$ latitude-longitude grid over the region  ($30^{\circ}$S-$30^{\circ}$N,$0^{\circ}$E-$360^{\circ}$E), resulting in $25\times144 = 3600$ grids. The dataset spans the time period between 1979-01-01 and 2016-12-31. El Ni$\tilde{n}$o or La Ni$\tilde{n}$a events are defined when the Oceanic Ni$\tilde{n}$o Index (ONI) exceed $\pm 0.5^\circ$C  for a period of five months or more. The ONI is defined as 3 month running mean of ERSST.v5 SST anomalies in the El Ni$\tilde{n}$o 3.4 region ($5^{\circ}$S-$5^{\circ}$N, $190^{\circ}$E-$240^{\circ}$E).

\subsection{\label{Methodology}Methodology}

In a climate system consisting of $N$ grids, the temperature of a grid $i$ at time $t$ within a time window of length $d$    
is defined as $S_i (t;T)$, where $T$ represents the central time of the time window. The average temperature of grid $i$ for the time window is calculated as
\begin{equation}
\bar{S_{i}^T}=\frac{1}{d}\sum_{t=1}^{d}S_{i}(t;T)\;.
\end{equation}

At the time $t$, the grid $i$ has a temperature fluctuation $\delta S_{i}(t;T)=S_{i}(t;T)-\bar{S_{i}^T}$. The normalized fluctuation $\delta \hat{S}_i (t;T) = \delta S_i (t;T) / \sqrt{\left\langle \left[ \delta S_{i}(t;T)\right]^{2} \right\rangle}$.
The correlation of temperature for the time window between grids $i$ and $j$ is defined as
\begin{equation}\label{cross1}
C_{ij}(T)=\left\langle \delta  \hat{S}_i (t;T)\cdot \delta  \hat{S}_j (t;T)\right\rangle\;,
\end{equation}
which is the Pearson correlation coefficient. Using $C_{ij}(T)$ as its elements, we obtain then a $N \times N$ correlation matrix $\bm{C}(T)$ with $N$ eigenvectors and eigenvalues. The eigenvector corresponding to eigenvalue $\lambda_{n}(T)$ is $\bm{b}_{n}(T)$ which satisfies the equation
\begin{equation}
\bm{C}(T)\cdot \bm{b}_{n}(T)=\lambda_{n}(T) \bm{b}_{n}(T),\; n=1,2,\ldots,N\;.
\end{equation}

The eigenvalues are numbered in the order $\lambda_{1}(T)\geq \lambda_{2}(T)\cdots\geq\lambda_{N}(T)\geq 0$. The eigenvectors follow the condition
\begin{equation}\label{evc}
\bm{b}_{n}(T)\cdot\bm{b}_{l}(T)=\sum_{i}b_{in}(T)b_{il}(T)=\delta_{nl}\;,
\end{equation}
where $\delta_{nl}=1$ at $n=l$ and $\delta_{nl}=0$ otherwise. All eigenvectors are normalized and orthogonal to each other.

Using the $n$-th eigenvector $\bm{b}_{n}(T)$, the $n$-th principal mode can be obtained as
\begin{equation} \label{pc}
a_{n}(t;T)=\sum_{i=1}^{N} b_{in}(T) \delta \hat{S}_i (t;T) \;.
\end{equation}
The correlation between principal modes $\bm{a}_{n}(T)$ and $\bm{a}_{l}(T)$ is
\begin{equation}
\label{cmo}
\left\langle \bm{a}_{n}(T)\cdot \bm{a}_{l}(T)\right\rangle=\frac{1}{d} \sum_{t=1}^{d} a_n(t;T) a_l(t;T) =\lambda_{n}(T)\delta_{nl}\;.
\end{equation}
So different principal modes are independent. The square amplitude of a principal mode is equal to the corresponding eigenvalue.

$\delta \hat{S}_i (t;T)$ can be considered also as the summation of contributions from all principal modes
\begin{equation} \label{pc1}
\delta \hat{S}_i (t;T) = \sum_{n=1}^{N}  b_{in}(T) a_{n}(t;T)  \;.
\end{equation}

Combining Eq. \ref{pc1} and Eq. \ref{cmo} into Eq. \ref{cross1}, the correlation between girds $i$ and $j$ can be written as  
\begin{equation}\label{cross2}
C_{ij}(T)=\sum_{n=1}^{N}  b_{in}(T) b_{jn}(T)\lambda_{n}(T) \;.
\end{equation}

To quantify the averaged correlation between a region A and a region B for the positive components of the n-th eigenvector, a "micro-correlation`` is introduced as
 \begin{equation}\label{cross3}
 C_{AB}^{n+}(T)=\frac{\sum_{i\in A}\sum_{j\in B}\Theta^+_{in}b_{in}(T)\Theta^+_{jn}b_{jn}(T)\lambda_{n}(T)}{\sum_{i\in A}\sum_{j\in B}} \;,
\end{equation}
where $\Theta^+_{in}=1$, when $b_{in}(T)>\Delta$, otherwise $\Theta^+_{in}=0$. $\Delta$ is a positive threshold to exclude weak and negative components. The definition of "micro-correlation`` for the negative components $C_{AB}^{n-}(T)$ is similar, only $\Theta^+_{in}$, $\Theta^+_{jn}$ change to $\Theta^-_{in}$, $\Theta^-_{in}$, where $\Theta^-_{in}=1$, when $b_{in}(T)<-\Delta$, otherwise $\Theta^-_{in}=0$.  

Furthermore, we can shift the time $T$ to get a series of correlation matrices $\bm{C}(T)$. Then the evolution of principal modes will be obtained. Note that different principal modes for different $T$ are not independent.  

\section{\label{sec:level3}Results}

\begin{figure}[htbp]
\centering
\includegraphics[scale=0.4]{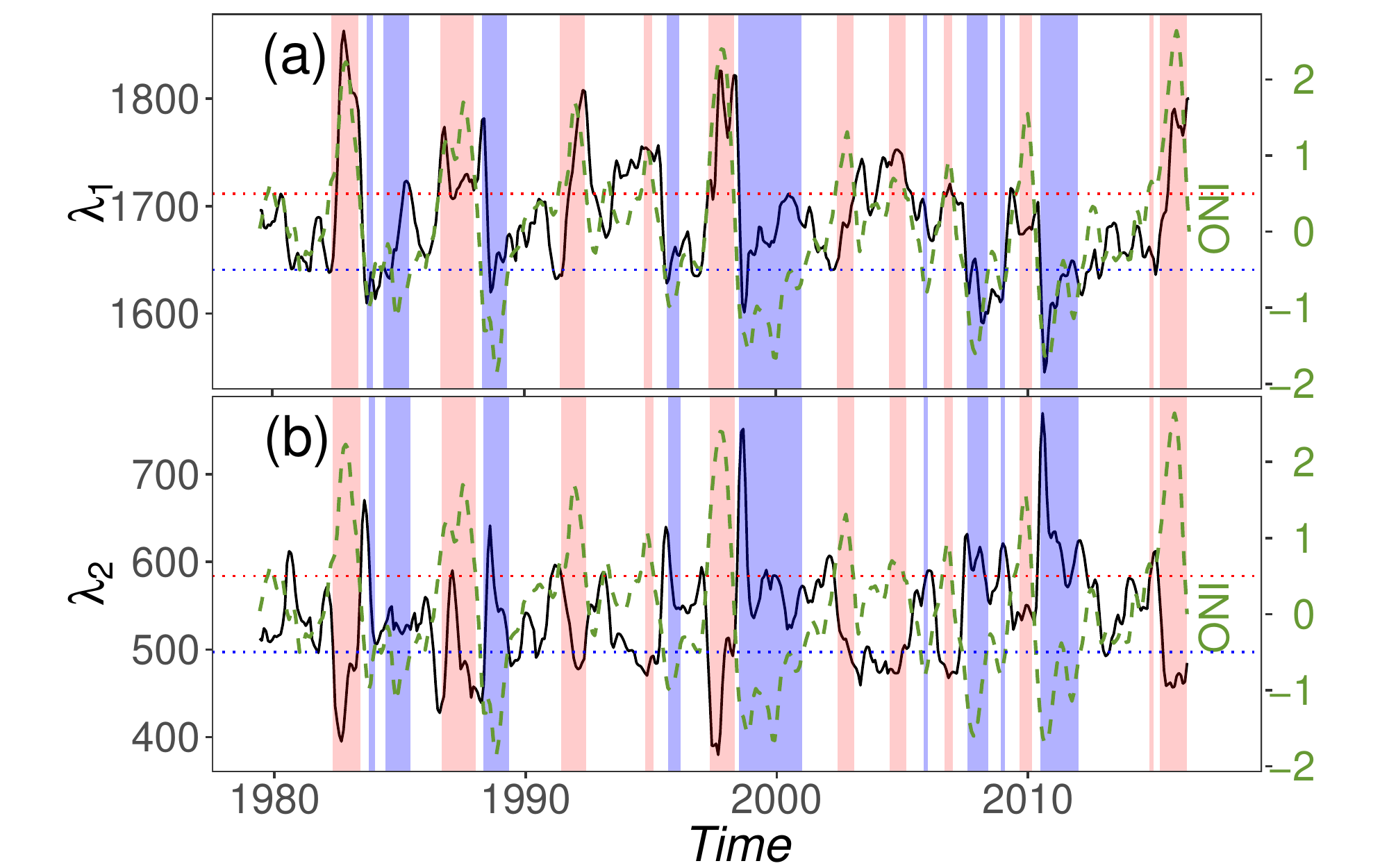}
\caption{ The first eigenvalue $\lambda_1$ (a) and the second eigenvalue $\lambda_2$ (b) as functions of time $T$ (black solid line, left scale) . The green dash line is the Oceanic Ni$\tilde{n}$o Index (ONI) (right scale). Red and blue shades represent respectively El Ni$\tilde{n}$o and La Ni$\tilde{n}$a periods.}\label{lambda12nino}
\end{figure}

Since the seasonal cycle is $365$ days, we take the length of time window $d=365$. For a time window with its center in the month $T$, we can obtain the correlation matrix $\bm{C}(T)$, its eigenvalues $\lambda_n (T)$, and the corresponding eigenvectors $\bm{b}_n (T)$. With $T$ shifted by one month each time, their evolution with time can be obtained. The first eigenvalues $\lambda_1 (T)$ and the second eigenvalues $\lambda_2 (T)$ with $T$ from 1979.06 to 2016.06 are shown in Fig. \ref{lambda12nino}(a) and (b). The ratio of the first eigenvalue $\lambda_1 (T)/\sum_{n}\lambda_n (T)$ is more than $40\%$. The second eigenvalue has the ratio $\lambda_2 (T)/\sum_{n}\lambda_n (T)$, which is more than $10\%$. So the temperature fluctuations in the region are dominated by these two principal modes.

We study at first the evolution of eigenvalues. After comparing $\lambda_1 (T)$ with the ONI, we find a strong positive relevance between them. The peaks of $\lambda_1 (T)$ correspond to the EI Ni$\tilde{n}$o events and its valleys correspond to the La Ni$\tilde{n}$a events (see Fig. \ref{lambda12nino}(a)). The overall correlation between $\lambda_1 (T)$ and the ONI is characterized by their Pearson correlation coefficient, which is equal to $0.65$. The Pearson correlation coefficient between $\lambda_2 (T)$ and the ONI is equal to $-0.62$. So $\lambda_2$ is negatively correlated to the ONI (see Fig. \ref{lambda12nino}(b)). Moreover, the Pearson correlation coefficient between $\lambda_1$ and $\lambda_2$ is calculated and equal to $-0.75$. This indicates that the temporal evolution of $\lambda_1 (T)$ is opposite to $\lambda_2 (T)$.

\begin{figure}[htbp]
\centering
\includegraphics[scale=0.4]{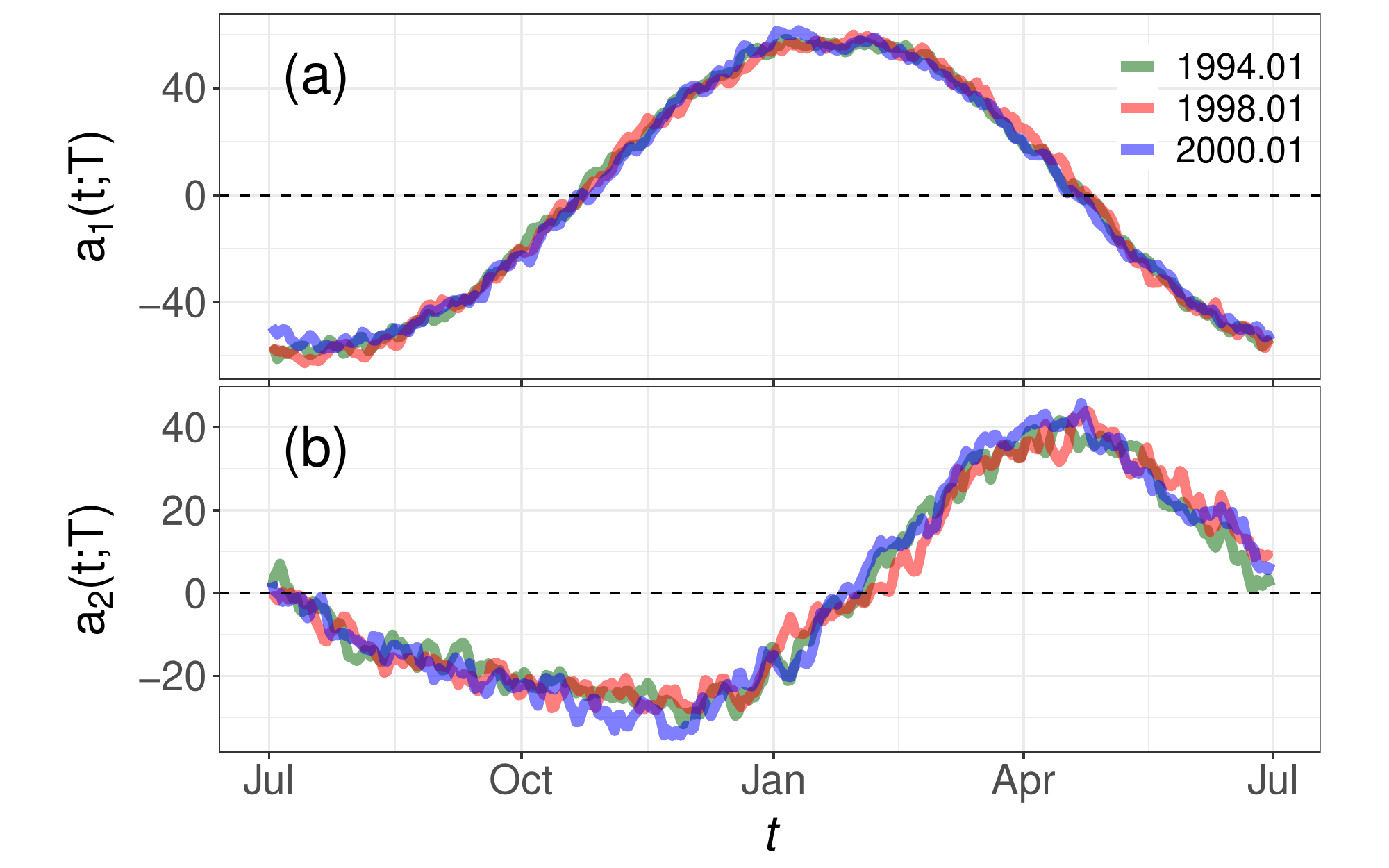}
\caption{(Color online) The (a) first and (b) second principal modes for the different $T$: 1994.01, 1998.01 and 2000.01.}\label{a12}
\end{figure}

From temperature fluctuations $\delta \hat{S}_i (t;T)$ and eigenvector $\bm{b}_n (T)$, we can get the intra-annual principal mode $\bm{a}_n(T)$ by Eq. \ref{pc}.
The principal mode $\bm{a}_n(T)$ is related to the eigenvalue by Eq. \ref{cmo}.
In Fig. \ref{a12}, the principal modes $a_1(t;T)$ and $a_2(t;T)$ are shown with respect $t$ at different $T$ corresponding to a normal time (1994.01), an El Ni$\tilde{n}$o time (1998.01), and a La Ni$\tilde{n}$a time (2000.01) respectively. A dominant seasonal oscillation in the first principal mode $a_1(t;T)$ is observed and presented in Fig. \ref{a12}(a). The maximum and minimum of $a_1(t;T)$ are located in mid February and mid August, which correspond respectively to winter and summer in the Northern Hemisphere. In the first principal mode $a_1(t;T)$, there are positive temperature fluctuations between mid November to next mid May and negative temperature fluctuations from mid May to mid November.

The second principal mode $a_2(t;T)$ is presented in Fig. \ref{a12}(b), where a seasonal oscillation is also observed. In comparison with $a_1(t;T)$, the seasonal oscillation of $a_2(t;T)$ has a shift. In relation to $a_2(t;T)$, there are positive temperature fluctuations from March to August and negative temperature fluctuations from August and next March.

The oscillation amplitude of $a_2(t;T)$ is smaller than that of $a_1(t;T)$. It is reasonable that the seasonal cycle plays the dominant role in $a_1(t;T)$ and $a_2(t;T)$, since the influences of Sun are the deciding factors to result in the fluctuations of surface air temperature. Besides the seasonal trend, there are only small differences of $a_1(t;T)$ and $a_2(t;T)$ for different $T$. Yet, these small differences can result in the significant correlations of $\lambda_1 (T)$ and $\lambda_2 (T)$ with the ONI.

We can express $\delta \hat{S}_i (t;T)$ by principal modes as Eq. \ref{pc1}. After taking into account the two dominant principal modes, we have
\begin{equation}
\delta \hat{S}_i (t;T) \approx b_{i1}(T) a_{1}(t;T) + b_{i2}(T) a_{2}(t;T).
\end{equation}

\begin{figure*}[htbp]
\centering
\includegraphics[scale=0.5]{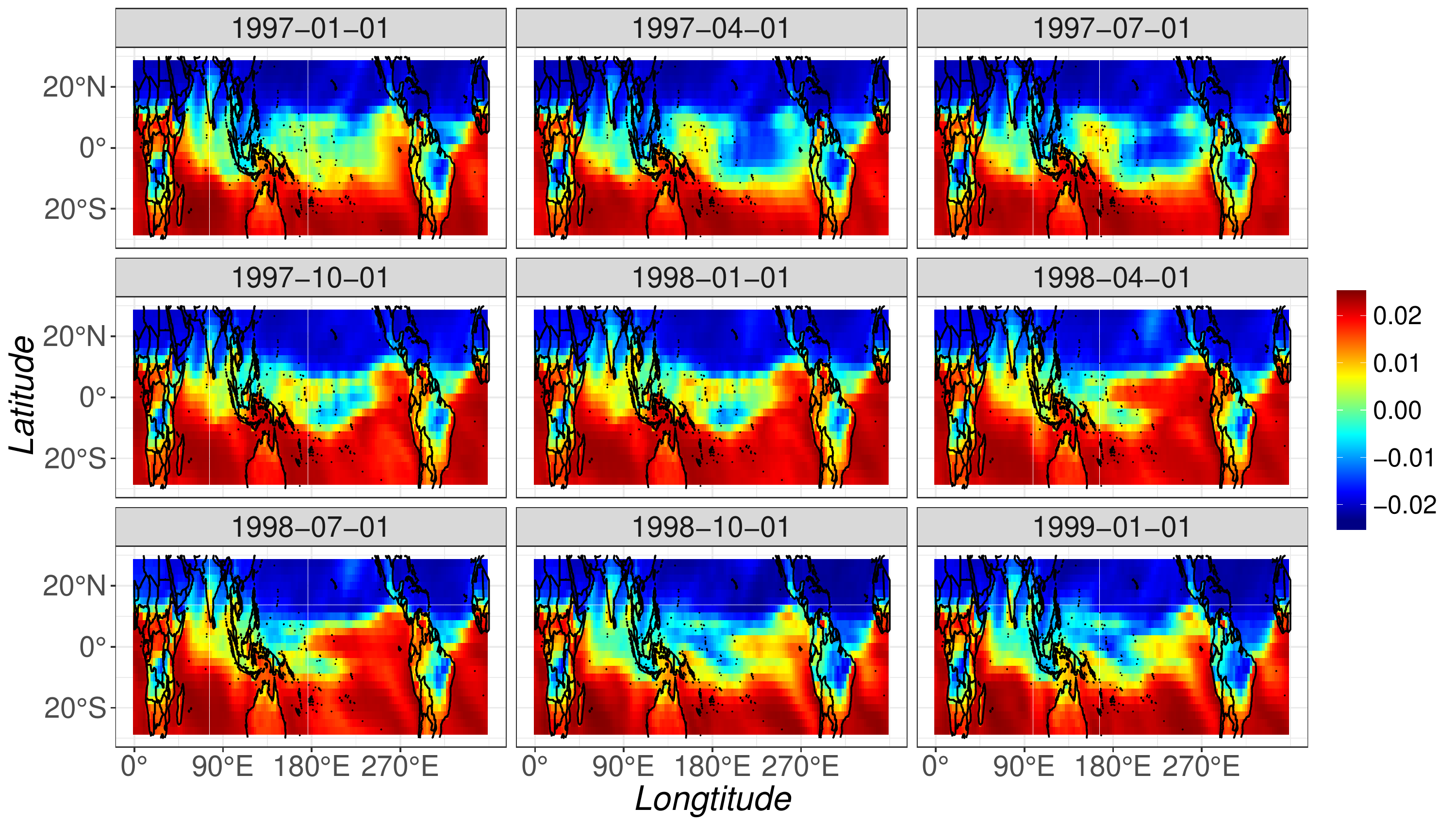}
\caption{(Color online) Spatial distributions of components of the first eigenvector $\bm{b}_{1}(T)$, where the time $T$ is from 1997.01 to 1999.01. They are the normal (1997.01--1997.04), El Ni$\tilde{n}$o (1997.05--1998.05), and  La Ni$\tilde{n}$a (1998.07--1999.01).}\label{eigenb1}
\end{figure*}

To understand further the physical features of intra-annual principle modes, we need to study the spatial distribution of $\bm{b}_n (T)$. The components $b_{i1}$ of the first eigenvector $\bm{b}_{1} (T)$ is depicted in Fig. \ref{eigenb1} for $T$ from 1997.01 to 1999.01, which spans the strong 97-98 El Ni$\tilde{n}$o event. The distributions of $b_{i1}$ are divided mainly into two large clusters. One is in the Northern Hemisphere and has negative components. Another one is in the Southern Hemispheres and has positive components. In addition, there are two small clusters with negative components and located in the rainforests of Congo (Africa) and Amazon (South America). The components of the interface between two large clusters are nearly zero. The interface exists around the Equator and varies greatly with time $T$.

In the running 3-month DJF of 1997, the ONI was $-0.49$ and increased to $0.74$ in the AMJ of 1997, which demonstrated the emergence of an El Ni$\tilde{n}$o event. This event lasted until to the MAM of 1998. With the further decrease of the ONI, a La Ni$\tilde{n}$a event appeared in the JJA of 1998 and lasted until 2001.

In the normal phase at the beginning, the El Ni$\tilde{n}$o 3.4 region ($5^{\circ}$S-$5^{\circ}$N, $190^{\circ}$E-$240^{\circ}$E) belongs to the interface and the corresponding $b_{i1}(T)$ are nearly zero (see Fig. \ref{eigenb1} for 1997.01). The temperature fluctuations in the region are nearly independent of the first principal mode. Later, the components $b_{i1}(T)$ in a part of the region become negative. The temperature fluctuations there are dominated by the first principal mode $\delta \hat{S}_i (t;T)\sim b_{i1}(T) a_{1}(t;T)$. The negative components (blue) in the El Ni$\tilde{n}$o 3.4 region (see Fig. \ref{eigenb1} for 1997.04 and 1997.07) result in the temperature fluctuations increases as $a_{1}(t;T)$ decreases from February to August (see Fig. \ref{a12}(a)) in comparison with the normal phase. Then an El Ni$\tilde{n}$o appeared, and the components in the El Ni$\tilde{n}$o region gradually change from negative to positive. We can see that the components $b_{i1}(T)$ in a part of the El Ni$\tilde{n}$o region become positive later (see Fig. \ref{eigenb1} for 1998.04 and 1998.07). These positive components (red) can contribute the increase of temperature fluctuations when $a_{1}(t;T)$ increases from August to the next February. After February, $a_1(t;T)$ decreases very fast so that the positive components $b_{i1}(T)$ in the region lead to a strong decrease of temperature and end the El Ni$\tilde{n}$o, which lasts until August. There is a La Ni$\tilde{n}$a now.

\begin{figure*}[htbp]
\centering
\includegraphics[scale=0.5]{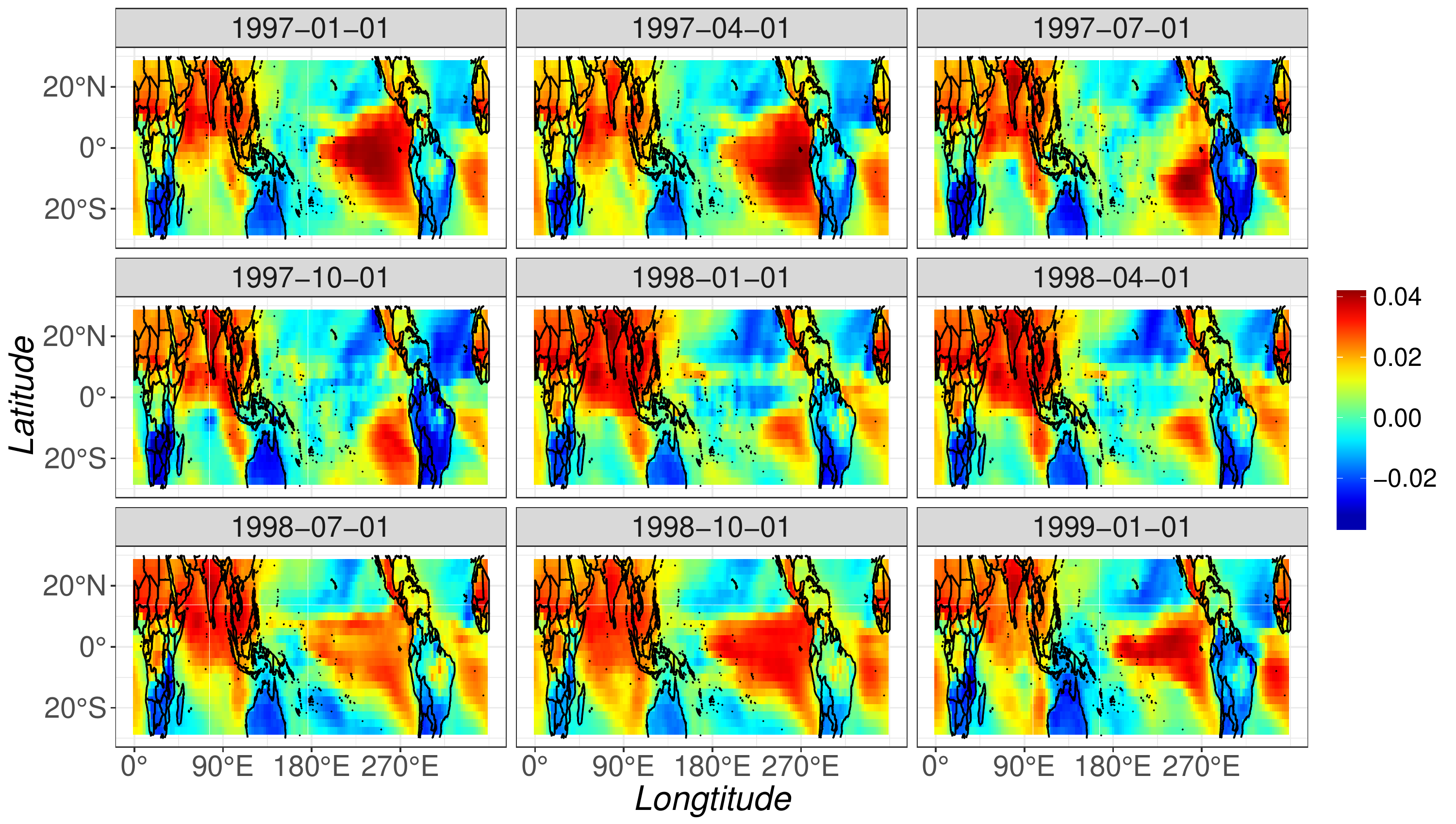}
\caption{(Color online) Same as Fig. \ref{eigenb1} but for the second eigenvector $\bm{b}_{2}(T)$.}\label{eigenb2}
\end{figure*}

The spatial distributions of $\bm{b}_{2}(T)$ are shown in Fig. \ref{eigenb2}. In the second principal mode, there are mainly two clusters with positive components and their temperature fluctuations are coupled. At the beginning of 1997, there is a large cluster with positive components in the Equatorial Eastern Pacific. With the emergence of El Ni$\tilde{n}$o from 1997.07 to 1998.04, this cluster becomes smaller and weaker. Even a cluster with negative components appears in the region. After the end of El Ni$\tilde{n}$o, there is a La Ni$\tilde{n}$a and a cluster with positive components appears again in the Eastern Pacific.

\begin{figure}[htbp]
\centering
\includegraphics[scale=0.4]{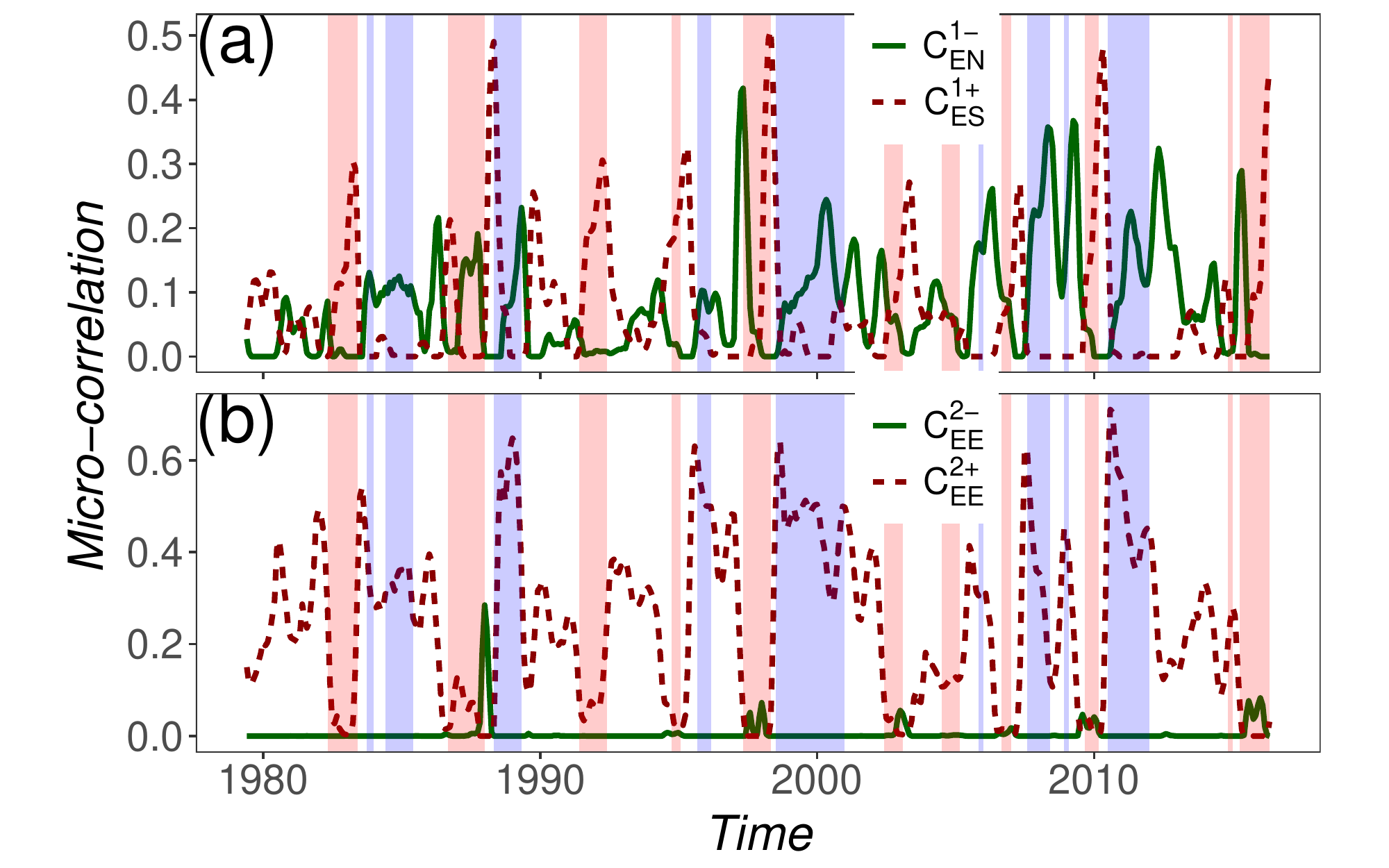}
\caption{(Color online) Micro-correlation for (a) the first eigenvector $\bm{b}_{1}$ and (b) the second eigenvector $\bm{b}_{2}$. Red and blue shades represent respectively El Ni$\tilde{n}$o and La Ni$\tilde{n}$a periods.}\label{I12nino}
\end{figure}

To further quantify the relationship between the ENSO and the spatial distributions of $\bm{b}_{n}$, we calculate the micro-correlation via Eq. \ref{cross3}. We take the threshold $\Delta$ as $\frac{1}{2\sqrt{N}}=0.008$. The micro-correlation between the El Ni$\tilde{n}$o 3.4 region and the north region ($5^{\circ}$N-$90^{\circ}$N, $0^{\circ}$E-$360^{\circ}$E) for the negative components of the first eigenvector can be obtained as $C_{EN}^{1-}$. Similarly we obtain the micro-correlation $C_{ES}^{1+}$ between the El Ni$\tilde{n}$o 3.4 region and the south region ($5^{\circ}$S-$90^{\circ}$S, $0^{\circ}$E-$360^{\circ}$E) for the positive components of $\bm{b}_{1}$. The micro-correlation $C_{EN}^{1-}$ (dashed line) and $C_{ES}^{1+}$ (solid line) are shown as functions of time $T$ in Fig. \ref{I12nino}(a). A large $C_{EN}^{1-}$ represents that the El Ni$\tilde{n}$o region is positively correlated with the Northern Hemisphere, as shown in Fig. \ref{eigenb1}. A large $C_{ES}^{1+}$ represents that the El Ni$\tilde{n}$o region is positively correlated with the Southern Hemisphere.
In normal phases, it can be seen in Fig. \ref{I12nino}(a) that both $C_{EN}^{1-}$ and $C_{ES}^{1+}$ are weak. Indeed, we observed that $C_{EN}^{1-}$ usually has a peak before the emergence of an El Ni$\tilde{n}$o event in Fig. \ref{I12nino}(a). Instead, $C_{ES}^{1+}$ has a sharp peak at the end of an El Ni$\tilde{n}$o event. 

Thus the evolution of the micro-correlation $C_{EN}^{1-}$ and $C_{ES}^{1+}$ could demonstrate that the ENSO phenomenon is related to the competition of the influences from the Northern and Southern Hemispheres. When the influences from both Hemispheres are in balance, the El Ni$\tilde{n}$o region belongs to the interface of $\bm{b}_1(T)$ so that its temperature fluctuations are independent of the Northern and Southern Hemispheres. There is a normal phase in the region. When the influences from the Northern Hemisphere dominate, the upper part of El Ni$\tilde{n}$o region is integrated into the northern cluster. The temperature increases in the El Ni$\tilde{n}$o region from February to August. Then, an El Ni$\tilde{n}$o event appears. If the influences from the Southern Hemisphere become dominant, the lower part of Ni$\tilde{n}$o region is integrated into the southern cluster. The temperature increases in the El Ni$\tilde{n}$o region until the next February, then fast decreases with the temperature of the Southern Hemisphere after February. We anticipate that the competition between the Northern and Southern Hemispheres in the first principal mode could be driven by the Hadley Circulation \cite{oort1996observed}, which is a general meridional ocean-atmospheric circulation.

Since the second eigenvector $\bm{b}_{2}$ is dominated by the cluster with positive components in the Equatorial Eastern Pacific (see Fig. \ref{eigenb2}), we calculate the micro-correlation for the El Ni$\tilde{n}$o 3.4 region itself like auto-correlation. Then we get $C_{EE}^{2-}$ and $C_{EE}^{2+}$ for the negative and positive  components respectively. Fig. \ref{I12nino}(b) shows that $C_{EE}^{2-}$ is always zero apart from during several El Ni$\tilde{n}$o events. The micro-correlation $C_{EE}^{2+}$ is strong during La Ni$\tilde{n}$a phases and weak during El Ni$\tilde{n}$o phases in Fig. \ref{I12nino}(b). It indicates that the strength of the cluster in the El Ni$\tilde{n}$o region is anti-correlated with the ONI in this region. It seems to be related to the Walker Circulation, which is a zonal ocean-atmosphere circulation in the Pacific \cite{Wang2017El}. The east-west surface temperature contrast reinforces an east-west air pressure difference across the Pacific basin, which in turn drives trade winds. During La Ni$\tilde{n}$a, the Walker Circulation becomes stronger. During El Ni$\tilde{n}$o instead, the trade winds become weak as atmospheric pressure rises in the western Pacific and falls in the eastern Pacific. Then the Bjerknes feedback operates in reverse, with weakened trade winds and SST warming tendencies along the equator \cite{Neelin2011Climate}. The Walker Circulation can provide the force to reshape the largest cluster in the Eastern Pacific of $\bm{b}_2$.

\begin{figure}[htbp]
\centering
\includegraphics[scale=0.4]{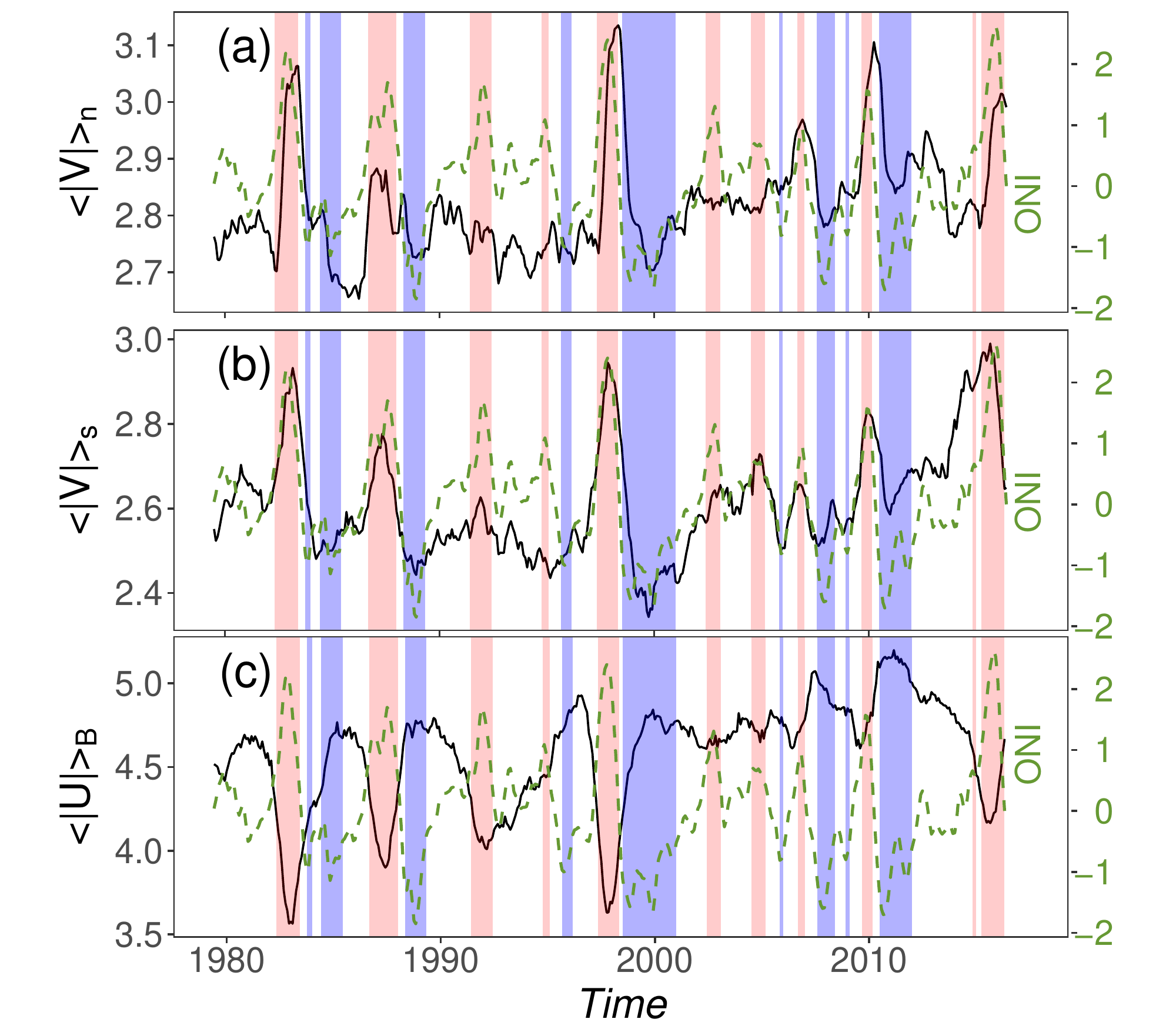}
\caption{The mean V wind strength (a) $\left\langle |V| \right\rangle_{s}$ and (b) $\left\langle |V| \right\rangle_{n}$, and (c) the mean U wind strength $\left\langle |U| \right\rangle_{B}$ as a function of time $T$.}\label{uv}
\end{figure}

We further prove our conjectures by studying the daily surface wind field (10 m). The wind is divided into meridional part $V$ and zonal part $U$. For the northern region ($0^{\circ}$-$30^{\circ}$N, $190^{\circ}$E-$270^{\circ}$E), the average magnitude $\left\langle |V| \right\rangle_{n}(T)$ is calculated and presented in Fig. \ref{uv}(a). We make the average for one year with its center in the month $T$.
The average magnitude of $V$ in the southern region ($0^{\circ}$-$30^{\circ}$S, $190^{\circ}$E-$270^{\circ}$E) is denoted as $\left\langle |V| \right\rangle_{s}(T)$. The results obtained are shown in Fig. \ref{uv}(b). We can see that El Ni$\tilde{n}$o events are accompanied usually by strong
$\left\langle |V| \right\rangle_{n,s}(T)$. In general, $\left\langle |V| \right\rangle_{n}(T)$ is larger than $\left\langle |V| \right\rangle_{s}(T)$. This result further demonstrates that the connection between the Equator and Northern (Southern) Hemisphere becomes stronger during El Ni$\tilde{n}$o in agreement with the results of the micro-correlation $C_{EN}^{1-}$ and $C_{ES}^{1+}$ in Fig. \ref{I12nino}(a).        

For the zonal part of wind, we calculate its average magnitude $\left\langle |U| \right\rangle_{B}(T)$ in the Ni$\tilde{n}$o 3 and 3.4 regions ($5^{\circ}$S-$5^{\circ}$N, $190^{\circ}$E-$270^{\circ}$E). The results obtained are depicted in Fig. \ref{uv}(c). During El Ni$\tilde{n}$o events, there are obvious decreases of $\left\langle |U| \right\rangle_{B}(T)$ as similar as $C_{EE}^{2+}$ in Fig. \ref{I12nino}(b).      

\section{\label{sec:level4}Conclusions}

Here we study the principal modes of surface air temperature in the region ($30^\circ$S-$30^\circ$N, $0^\circ$E-$360^\circ$E) within one year in relation to the El Ni$\tilde{n}$o/Southern Oscillation. They are dominated by the two largest intra-annual principal modes, which are obtained by the eigen-decomposition method. The temporal evolution of the principal modes is investigated from 1979-01-01 to 2016-12-31. Their eigenvalues $\lambda_1$ and $\lambda_2$ response oppositely to the ENSO variability. The Pearson correlation coefficients between the two eigenvalues and the ONI are equal to $0.65$ and $-0.62$, respectively.

Both principal modes exhibit significant seasonal oscillations. We propose an evolutionary mechanism of El Ni$\tilde{n}$o/Sothern Oscillation based on the temporal evolution of the spatial distribution of two principal modes. The first principal mode $a_n(t;T)$ decreases as $t$ from February to August, and increases from August to the next February. In normal phases, the El Ni$\tilde{n}$o region has weak coupling with the first principal mode and its temperature fluctuations are dominated by the second principal mode. When the coupling becomes strong between the El Ni$\tilde{n}$o region and the Northern Hemisphere between February and August, an El Ni$\tilde{n}$o event will occur with a high probability. As the evolution of the first eigenvector, the coupling changes to be strong between the El Ni$\tilde{n}$o region and the Southern Hemisphere. It causes that the temperature increases in the El Ni$\tilde{n}$o region, then fast decreases after the next February. We introduce the micro-correlation to quantify the correlations between the El Ni$\tilde{n}$o region and the Northern (Southern) Hemisphere for the principal mode. On the other hand, the coupling between the El Ni$\tilde{n}$o region and the second principal mode is strong in normal or La Ni$\tilde{n}$a phases in response to the normal or strong Walker Circulations. With the emergence of an El Ni$\tilde{n}$o event, this coupling becomes much smaller so that there is a much weaker Walker Circulation.

We suggest that the first and second intra-annual principal modes are related to the meridional and zonal circulations respectively. This is partly demonstrated by the mean meridional and zonal surface wind field ($10$ m). During El Ni$\tilde{n}$o events, there are stronger mean meridional wind and weaker mean zonal wind. The meridional circulations can drive the competition between the influences from the Northern and Southern Hemispheres that could determine the El Ni$\tilde{n}$o region to be in a normal, El Ni$\tilde{n}$o, and La Ni$\tilde{n}$a phases.

\begin{acknowledgments}
We thank  the financial support by the National Natural Science Foundation of China (Grant Nos. 61573173) and Key Research Program of Frontier Sciences, CAS (Grant No. QYZD-SSW-SYS019). We also acknowledge the computational resources provided by HPC Cluster of ITP-CAS. Data in support of this work can be found through the ERA‐-Interim reanalysis of the European Center for Medium‐-Range Weather Forecasts (https://doi.org/ 10.1002/qj.828).
\end{acknowledgments}

%

\end{document}